\documentclass[]{aa}
\usepackage{graphicx}
\usepackage{txfonts}

\title{\textsl{INTEGRAL} observation of the high-mass X-ray transient  
V\,0332+53 during the 2005 outburst decline}
\titlerunning{V\,0332+53 during the 2005 outburst decline}
\author{N.~Mowlavi \inst{1,2},
		I.~Kreykenbohm \inst{1,3},
		S.~E.~Shaw \inst{1,4},
		K.~Pottschmidt \inst{5},
		J.~Wilms\inst{6},
		J.~Rodriguez \inst{7,1},
		N.~Produit \inst{1},
		S.~Soldi \inst{1,2},
		S.~Larsson \inst{8}
	\and 	P.~Dubath \inst{1,2}
	}
\authorrunning{N. Mowlavi, I. Kreykenbohm, S. E. Shaw et al.}
\institute{\textsl{INTEGRAL} Science Data Centre, ch. d'\'Ecogia, 1290 Versoix, Switzerland
\and
Observatoire de Gen\`eve, ch. des Maillettes, 1290 Versoix, Switzerland
\and
Institut f\"ur Astronomie und Astrophysik - Astronomie, Sand 1, 72076 T\"ubingen, Germany
\and
School of Physics and Astronomy, University of Southampton, SO171BJ, UK
\and
Center for Astrophysics and Space Sciences, University of California at San Diego, La Jolla, CA 92093-0424, U.S.A.
\and
Department of Physics, University of Warwick, Coventry, CV4 7AL, UK
\and
CEA Saclay, DSM/DAPNIA/SAp(CNRS UMR 7158 AIM), 91191Gif Sur Yvette,  
France
\and
Stockholm Observatory, Alba Nova, 106 91 Stockholm, Sweden}

\date{Received September 21, 2005; Accepted December 14, 2005}

\abstract{The decline of the high mass X-ray transient V\,0332+53  during the Dec. 2004 to Feb. 2005 outburst
	is analysed from the data recorded by \textsl{INTEGRAL}.
		The flux is shown to decrease exponentially until 2005 Feb. 10, with a decay time scale
	of $\sim$30~days above 20\,keV and $\sim$20~days at lower energies, and to decrease linearly thereafter.\\
		The energy spectrum is well modelled throughout the decay by a  power law with a
		folding energy of $\sim$7.5\,keV, and with two cyclotron absorption features.
	The folding energy does not vary significantly over the decay, but the spectrum becomes harder
	with time.
	Most importantly, we show that the parameters describing the fundamental cyclotron line around 27\,keV do
     vary with time:
	its energy and depth increase (by about 17\% for the energy in $\sim$6 weeks), while its width decreases.
	These changes of the cyclotron line parameters are interpreted as resulting from a change in the extent of the
	cyclotron scattering region.\\
		Two quasi-periodic oscillations are also observed at various times  during the observations,
		one at 0.05\,Hz  and another one near the pulsation frequency  around  0.23\,Hz.

}

\begin{document}
\maketitle

\section{Introduction}

The recurrent X-ray transient V\,0332+53 experienced an X-ray outburst from December 2004 to February 2005.
The outburst was predicted one year earlier from an optical  brightening of the massive O8-9Ve companion star, BQ Cam
(Goranskij \& Barsukova \cite{Goranskij_Barsukova04})
and was first detected by the Rossi X-ray Timing Explorer ({\it RXTE}) in November 2005 (Swank et al. \cite{Swank_etal04}).
The INTernational Gamma-Ray Astrophysics Laboratory (\textsl{INTEGRAL}) started to observe the source
on 2005 Jan. 6 (Kreykenbohm et al. \cite{Kreykenbohm_etal05}, hereafter Paper~I)
and monitored its X-ray activity until the end of February.

The observation of this fourth detected outburst of V\,0332+53 in thirty years
revealed the presence of three cyclotron lines
(Coburn et al. \cite{Coburn_etal05}, Paper~I, Pottschmidt et al. \cite{Pottschmidt_etal05}).
The fundamental has an energy of 28\,keV, and the first and second harmonics are at 50\,keV and 71\,keV, respectively.
The magnetic field strength is $2.7 \times 10^{12}$\,G (Paper~I, Pottschmidt et al. \cite{Pottschmidt_etal05}).
The spectrum is otherwise described by a power law with a high-energy cut-off, typical of the spectra of
X-ray pulsars.
A new Quasi Periodic Oscillation (QPO) around the 0.23\,Hz spin frequency of the neutron star has also been
reported in the {\it RXTE} data (Qu et al. \cite{Qu_etal05}), in addition to the 0.05\,Hz QPO
discovered by Takeshima et al. (\cite{Takeshima_etal94}) in the {\it Ginga} data of the 1989 outburst.

In Paper I, we presented the \textsl{INTEGRAL} observation of V\,0332+53 around the outburst peak.
This paper presents the evolution of V\,0332+53 during its decline  phase.
The observations and data analysis method are described in Sects \ref{Sect:observations} and \ref{Sect:data_analysis}, respectively.
We then present in Sect.~\ref{Sect:results} the evolution during the  decline of the fluxes in several energy bands,
of the energy spectrum, and of the variability spectrum of V\,0332+53.
The results are then discussed in Sect.~\ref{Sect:discussion}.

\section{Observations}
\label{Sect:observations}

\begin{table*}
\caption{Summary of observations. From first to last column:
          (1) Revolution and pointing numbers;
          (2) Earth date of the start of the observation in UTC / IJD  (\textsl{INTEGRAL} Julian Day = MJD\,$-$\,51544) / MJD (Modified Julian Day);
          (3) Earth date of the end of the observation in MJD;
          (4) Individual pointing duration;
          (5) Total IBIS observation time;
          (6) Sum of IBIS Good Time Intervals;
          (7) Total JEM-X1 observation time;
          (8) Sum of JEM-X1 Good Time Intervals.
          }
\label{Tab:observations}
\centering
\renewcommand{\footnoterule}{}  
\begin{tabular}{l c c c r r r r}
   \hline
                             &                                      		     &                    & pointing  & \multicolumn{2}{c}{IBIS}       & \multicolumn{2}{c}{JEM-X1} \\
   Revolution /    &                              Start date 			&   End date  & duration & Exposure & $\Sigma$ GTIs & Exposure & $\Sigma$ GTIs   \\
   \multicolumn{1}{r}{pointings}& (UTC / IJD / MJD)	&    (MJD)     &  (sec)       & (sec)~        &        (sec)~         &   (sec)~      &        (sec)~           \\
   \noalign{\smallskip}
   \hline
   272 / 75         & 2005-01-06T08:48  / 1832.36724 / 53376.36724  &53376.76044 &  42184    & 42184 &          30308 &     36896 &           36446 \\
   273 / 66 - 81  & 2005-01-08T22:18 / 1834.92997 / 53378.92997  &53379.65749 &  3500 & 60953 &          59586 &     60953 &            56693 \\
   274 / 04 - 09  & 2005-01-10T03:20 / 1836.13991 / 53380.13991  &53380.39113 &  3500 & 21119 &          20901 &     21119 &            19801 \\
    \noalign{\smallskip}
   278 / 48 - 74  & 2005-01-23T16:39 / 1849.69471 / 53393.69471  &53394.66610 &  2858 & 80901 &          58411 &     75190 &            73078 \\
    \noalign{\smallskip}
   284 / 04 - 45  & 2005-02-09T01:15 / 1866.05280 / 53410.05280  &53410.86456 &  3572 & 149086 &        147739 &  149086 &           147469 \\
   285 / 02 - 08  & 2005-02-12T01:02 / 1869.04402 / 53413.04402  &53413.23174 &  2200 & 15517 &           15349 &    15517 &            14394 \\
   286 / 18 - 24  & 2005-02-15T10:41 / 1872.44588 / 53416.44588  &53416.63218 &  2200 & 15403 &           15093 &    15403 &            15185 \\
   287 / 02 - 08  & 2005-02-18T00:37 / 1875.02612 / 53419.02612  &53419.23077 &  2200 & 16988 &           16628 &    16988 &            15809 \\
   288 / 02 - 08  & 2005-02-21T00:25 / 1878.01820 / 53422.01820  &53422.25899 &  2200 & 20101 &           19226 &    20101 &            18854 \\

   \hline
\end{tabular}
\end{table*}

The \textsl{INTEGRAL} satellite carries three high energy instruments,
the imager IBIS (operating in the range 15 keV to 10 MeV),
the spectrometer SPI (20 keV to 8 MeV) and
two identical X-ray instruments JEM-X1 and JEM-X2 (3 to 35 keV)  operational in turn
(see Winkler et al. \cite{Winkler_etal03} for more information on the  
\textsl{INTEGRAL} mission and instruments).
IBIS is further composed of two modules, ISGRI operational down to 15\,keV and PICsIT above 200\,keV.

\textsl{INTEGRAL} observed V\,0332+53 from 2005 Jan. 6 to Feb. 21.
Three main observation sets
were scheduled, in revolutions 272-274 ($\sim$120\,ksec), 278 ($\sim$80\,ksec) and 284 ($\sim$150\,ksec),
followed by four $\sim$15\,ksec observations from revolution 285 to  
288 to monitor the tail of the outburst decay.
The observations are summarized in Table~\ref{Tab:observations}.
Except for the observation in revolution 272, which
was performed in a staring mode, all observations were made following  
a hexagonal dithering pattern.
The hexagonal dithering pattern consists of one pointing (stable  
spacecraft attitude) with
V\,0332+53 positioned at the center of the field of view of the  
instruments, and six
pointings with V\,0332+53 off-axis in steps of 2 degrees in a  
hexagonal pattern.
The integration time for each pointing may vary from one observation  
to the next and is given in
Table~\ref{Tab:observations}.

The V\,0332+53 observations were affected by solar activity which  
impacted on the quality of the data.
In particular, a strong solar flare lasting from Jan. 20 to 23 lead  
to a split of the IBIS data of revolution 278
into 7760 small Good Time Intervals (GTIs) of only few seconds each.
The JEM-X GTIs were greater than 97\% of the observation time for  
that revolution, and in general greater than 90\% for all revolutions  
(see Table~\ref{Tab:observations}).

\section{Data analysis}
\label{Sect:data_analysis}

We analyse all the public data collected by the ISGRI and the JEM-X1  
instruments on
board of  \textsl{INTEGRAL}.
PICsIT is not useful for our analysis because the spectrum of V0332+53 falls off above 100\,keV,
and SPI data are not used either since its sensitivity in the 20 to  
100 keV range is less favorable than ISGRI (Paper I).
Pointings close to perigee passage times are further discarded from  
the analysis as they are contaminated by
the Earth radiation belt.

The ISGRI and JEM-X data are processed with version 5.0 of the  
Offline Scientific Analysis (OSA) package delivered by
the \textsl{INTEGRAL} Science Data Center (http://isdc.unige.ch)  
using default settings.
Images and lightcurves are produced in the 3--5\,keV, 5--10\,keV and  
10--15\,keV energy bands for JEM-X,
and in the 20--30\,keV, 30--40\,keV and 40--60\,keV bands for ISGRI.

Contamination from other sources is not a problem as
V\,0332+53 is the only bright X-ray source in the ISGRI field of view.
Systematic errors are however known to affect the data analysis.
In order to estimate the amplitude of those systematic errors,
and to calibrate the results, an analysis of the Crab was performed
with OSA parameters identical to those used for the analysis of V\,0332+53.
Crab public data are taken from revolution 239, pointings 3 to 27
(Modified Julian Day MJD=53275.5289 to 53276.1326),
during which the Crab is located within 3 degrees from the center of  
the field of view.
The resulting mean Crab fluxes extracted from the images in the 3--5,  
5--10, 10--15\,keV bands of JEM-X are
43.5 cps (14\% error), 43.2 cps (10\%) and 16.4 cps (12\%),  
respectively.
The quoted errors represent the maximum deviations measured from the  
mean fluxes over the 25 pointings.
The Crab fluxes in the 20--30, 30--40 and 40--60\,keV energy bands of  
ISGRI are 78.8 cps (13\%), 39.4 cps (7\%)
and 45.9 cps (4\%), respectively.

\section{Results}
\label{Sect:results}

\subsection{Fluxes}
\label{Sect:flux}

\begin{figure}
    \centering
    \includegraphics[width=8.7cm]{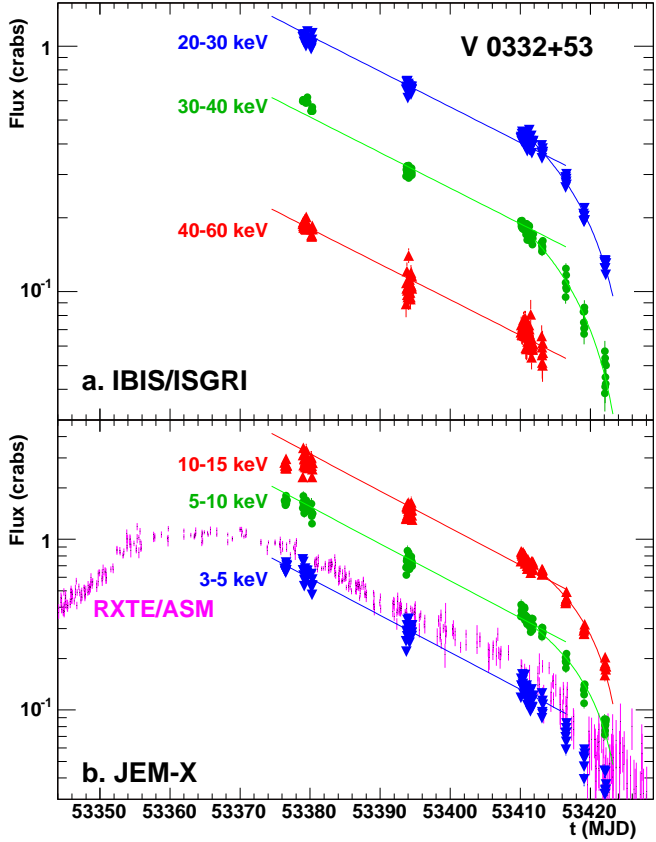}
    \caption{{\bf a.} Pointing averaged ISGRI flux evolution of V\,0332+53
              as a function of time in the different energy bands as  
labelled in the figure.
              The fluxes have been normalised to Crab units.
              The error bars displayed in the figure represent only  
the statistical errors; neither do they take into account the  
uncertainties in the Crab fluxes in the respective energy bands.
              The solid curves are exponential (from MJD=53376 to  
53412 with $\tau$=30 days) and linear
              (from MJD=53412 to 53422) decays superimposed on the data.
             {\bf b.} Same as {\bf a}, but for JEM-X fluxes and with  
an exponential decay of $\tau$=20 days
             for the solid curves.
             The 2--12\,keV RXTE/ASM lightcurve is also shown, with a  
normalisation factor of 77 cps for 1 Crab.
            }
    \label{Fig:flux}
\end{figure}

\begin{figure}
    \centering
    \includegraphics[width=8.7cm]{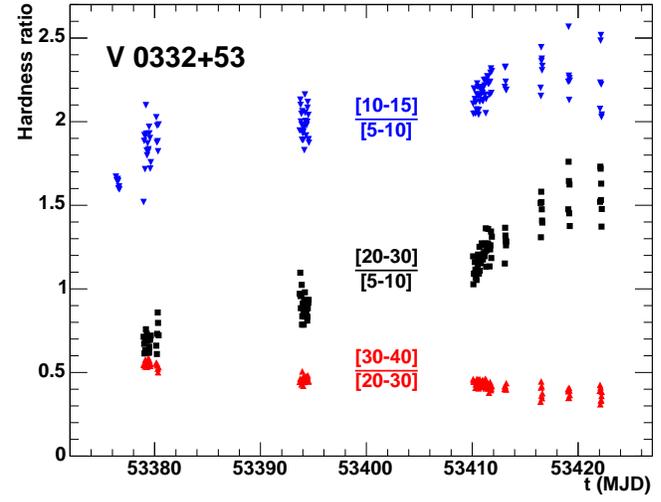}
    \caption{Hardness ratios computed from the pointing-averaged (and  
Crab normalized) fluxes
             displayed in Fig.~\ref{Fig:flux}.
             Upper triangles: ISGRI [30--40\,keV]/[20--30\,keV];
             lower triangles: JEM-X [10--15\,keV]/[5--10\,keV];
             squares: ISGRI [20--30\,keV] / JEM-X [5--10\,keV].
             }
    \label{Fig:hr}
\end{figure}

The fluxes of V\,0332+53 in the ISGRI and JEM-X energy bands are  
plotted in Fig.~\ref{Fig:flux}
as a function of time. They are normalised to the Crab fluxes in the  
respective bands.

The decline phase can be described by an exponential decay up to MJD$\simeq$53412,
followed by a linear decrease. The exponential decay is less
rapid at the ISGRI energies than at the JEM-X energies.
 From a simple estimation, the exponential decay time is about 30  
days above 20\,keV (solid curves in Fig.~\ref{Fig:flux}a)
and about 20 days below 15\,keV (solid curves in Fig.~\ref{Fig:flux}b).
The exponential decay is also confirmed by the 2--12\,keV lightcurve  
recorded by the All Sky Monitor (ASM) on board of RXTE,
plotted in Fig.~\ref{Fig:flux}b for comparison with the JEM-X  
lightcurves
(see Levine et al. \cite{Levine_etal96} for more information on the  
{\it RXTE} satellite).

The evolution of the ISGRI and JEM-X hardness ratios are shown in  
Fig.~\ref{Fig:hr}.
Above 20\,keV the spectrum of V\,0332+53 softens with time, with the  
[30--40]\,keV\,/\,[20--30]\,keV
ratio decreasing by about 36\% from an average value of $\sim$0.55 in  
the first observation to $\sim$0.35 in the last observations.
On the other hand, below 15\,keV, the spectrum becomes harder, with  
the [10--15]\,keV\,/\,[5--10]\,keV
ratio increasing by about 28\% from an average value of $\sim$1.8 to $\sim$2.3 during the decline.
As will be seen in Sect.~\ref{Sect:spectrum}, the opposite behaviour  
of the hardness ratios above 20\,keV
and below 15\,keV are understood from the evolution of the spectral  
shape of V\,0332+53 during the decline.

The ISGRI/JEM-X hardness ratio ([20--30]\,keV\,/\,[5--15]\,keV) also  
evolves with time, increasing from an average of
$\sim$0.7 to $\sim$1.5 during the decline.
This results from the increase of the spectral power law index  
presented in the next section.

\subsection{Energy spectra}
\label{Sect:spectrum}

\begin{table*}
\caption{Fitting parameters. From first to last column:
          (1) revolution number (see Table~\ref{Tab:observations});
          (2) normalisation factor;
          (3) powerlaw index;
          (4) folding energy (\textsf{cutoffpl} in XSPEC);
          (5) energy of the fundamental cyclotron line;
          (6) width of the line;
          (7) optical depth at the line center;
          (8) flux in the 2--10\,keV band;
          (9) flux in the 20--40\,keV band.
          }
\label{Tab:spefit}
\centering
\renewcommand{\footnoterule}{}  
\begin{tabular}{l c c cc cccccc cc}
   \hline
  Rev. & $A$		& $\Gamma$	& $E_{\rm{fold}}$	& $E_{\rm{cycl,1}}$	&	$\sigma_{1}$	& $\tau_{1}$	& $E_{\rm{cycl,2}}$	&	$\sigma_{2}$	& $\tau_{2}$	& 2--10\,keV				 & 20--40\,keV\\
  	 &		&			& keV		     &	keV		              &  keV                      &                     &  keV                          & keV                        &                    &  erg/cm$^{2}$/s & erg/cm$^{2}$/s\\
   \hline
   273 & $0.458^{+0.004}_{-0.002}$& $-0.174^{+0.005}_{-0.004}$	&  $7.04^{+0.01}_{-0.02}$	& $27.53^{+0.04}_{-0.04}$ 	& $4.70^{+0.03}_{-0.03}$	& $1.80^{+0.01}_{-0.01}$
   	& $51.5^{+0.3}_{-0.1}$	& $9.6^{+0.1}_{-0.2}$	& $1.82^{+0.02}_{-0.02}$	&$1.89\times10^{-8}$ & $7.31\times10^{-9}$ \\
   274 &   $0.35^{+0.01}_{-0.00}$	&   $-0.26^{+0.02}_{-0.02}$		& $6.9^{+0.1}_{-0.1}$		&   $27.7^{+0.1}_{-0.1}$ 		&   $4.7^{+0.1}_{-0.1}$		& $1.78^{+0.01}_{-0.04}$
   	& $55.0^{+1.3}_{-0.2}$	& $11.3^{+1.3}_{-0.6}$	& $2.0^{+0.1}_{-0.1} 
$	& $1.66\times10^{-8}$ & $6.92\times10^{-9}$ \\
    \noalign{\smallskip}
   278 &	  $0.17^{+0.01}_{-0.00}$	&   $-0.24^{+0.02}_{-0.07}$		&    $7.5^{+0.3}_{-0.3}$		&   $28.8^{+0.1}_{-0.1}$ 		&   $4.8^{+0.1}_{-0.1} 
$		& $1.92^{+0.04}_{-0.03}$
   	& $54.6^{+1.1}_{-0.8}$	& $9.7^{+0.7}_{-0.5}$	& $2.6^{+0.3}_{-0.1}$	& $8.53\times10^{-9}$ & $4.37\times10^{-9}$ \\
    \noalign{\smallskip}
   284 &	  $0.06^{+0.00}_{-0.00}$	&   $-0.41^{+0.00}_{-0.01}$		&    $7.4^{+0.0}_{-0.0}$		&   $29.3^{+0.0}_{-0.0}$ 		&   $4.3^{+0.0}_{-0.0} 
$		& $1.96^{+0.02}_{-0.02}$
   	& $51.0$ fix	& $7.4^{+0.1}_{-0.1}$	& $2.0^{+0.1}_{-0.1}$	& $3.94\times10^{-9}$ & $2.69\times10^{-9}$ \\
   285 &	  $0.04^{+0.00}_{-0.00}$	&   $-0.53^{+0.01}_{-0.03}$		&    $7.2^{+0.0}_{-0.3}$		&   $29.4^{+0.1}_{-0.1}$ 		&   $4.4^{+0.0}_{-0.1} 
$		& $1.96^{+0.04}_{-0.04}$
   	& $51.0$ fix	& $7.9^{+0.2}_{-0.2}$	& $2.1^{+0.1}_{-0.1}$	& $3.15\times10^{-9}$ & $2.41\times10^{-9}$ \\
   286 &	  $0.03^{+0.00}_{-0.00}$	&   $-0.44^{+0.05}_{-0.06}$		&    $8.0^{+0.2}_{-0.4}$		&   $29.5^{+0.3}_{-0.1}$ 		&   $4.5^{+0.1}_{-0.1} 
$		& $2.00^{+0.06}_{-0.05}$
   	& $51.0$ fix	& $7.9^{+0.2}_{-0.8}$	& $2.4^{+0.2}_{-0.1}$	& $2.19\times10^{-9}$ & $1.80\times10^{-9}$ \\
   287 &	  $0.02^{+0.00}_{-0.00}$	&   $-0.41^{+0.05}_{-0.08}$		&    $8.3^{+0.3}_{-0.5}$		&   $29.4^{+0.3}_{-0.1}$ 		&   $4.1^{+0.2}_{-0.1} 
$		& $2.11^{+0.08}_{-0.07}$
   	& $51.0$ fix	& $8.8^{+0.4}_{-1.0}$	& $2.7^{+0.3}_{-0.2}$	& $1.51\times10^{-9}$ & $1.28\times10^{-9}$ \\
   \hline
\end{tabular}
\end{table*}

\begin{figure}
    \centering
    \includegraphics[width=8.7cm]{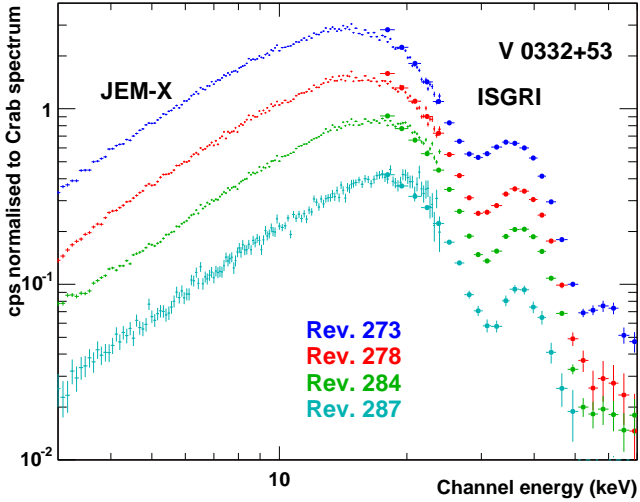}
    \caption{JEM-X and ISGRI spectra integrated, from upper to lower curves,
                  during revolutions 273, 278, 284 and 287, respectively.
                  The spectra have been normalised to the Crab spectrum of revolution 239 (see text for details).
            }
    \label{Fig:spe}
\end{figure}

The JEM-X and ISGRI spectra of V\,0332+53, normalised to the Crab  
spectrum,
are shown in Fig.~\ref{Fig:spe} at four different times during the decline of the source.
The maximum flux relative to the Crab is recorded between 10 and 20\,keV,
at an energy $E_{\rm max}$ which increases with time.
Fitting gaussian curves to the JEM-X spectra in a 10\,keV width range around the maximum leads
to $E_{\rm max}=14.6$, 15.4, 16.8 and 18.0\,keV for revolutions 273, 278, 284 and 287, respectively.
This shift of $E_{\rm max}$ explains the hardening (softening) at low (high) energies noticed
from the JEM-X (ISGRI) hardness ratios presented in Sect.~\ref{Sect:flux}.

The spectra of accreting X-ray pulsars include contributions from at least
an accretion column and from the hotspots at the surface of the neutron star.
In the absence of a consistent model, various empirical models are commonly used
in the literature to fit such spectra  (see for example Kreykenbohm et al. \cite{Kreykenbohm_etal99} for a discussion).
In this study, we use the simple ``\textsf{cutoffpl}'' model of the form
\begin{equation}
	F(E) = A \cdot E^{-\Gamma} \cdot e^{-E/E_{\rm{fold}}}
\label{Eq:cutoffpl}
\end{equation}
to fit the combined ISGRI and JEM-X data, to which we add two  
cyclotron lines (see below).
A multiplicative constant is applied to allow for the different  
normalisations of the two instruments.
Furthermore a systematic error of 2\% is applied to account for the  
uncertainties in the response matrices of
JEM-X and ISGRI.
No photoelectric absorption is required at low energies, compatible  
with the fact
that we analyse the JEM-X data only above 4\,keV; the addition of a  
Fe K$\alpha$ fluorescence line does not improve the fits either.

To fit the cyclotron lines we use Gaussian optical depth profiles of the form
\begin{equation}
	G_{i}(E) = \tau_{i}
	               \cdot e^{ -\frac{1}{2}\left(\frac{E-E_{{\rm cycl,}i}} 
{\sigma_i}\right)^2 }
\label{Eq:gabs}
\end{equation}
where $E_{\rm cycl}$ is the line energy, $\sigma$ the line width, and  
$\tau$ the optical depth
at the line center (see Kreykenbohm et al. \cite{Kreykenbohm_etal04},  
Pottschmidt et al. \cite{Pottschmidt_etal05}).
The index $i$ identifies the line, i.e $i$=1 for the fundamental at 28\,keV and $i$=2 for the first harmonic at 50\,keV.
The second harmonic at 71\,keV (Paper~I)  is not modelled due to  
insufficient statistics for revolution averaged spectra.
The overall spectral model is then given by
\begin{equation}
f(E) = F(E) \cdot e^{-G_{1}(E)} \cdot e^{-G_{2}(E)}
\label{Eq:model}
\end{equation}
where $F(E)$ is given by Eq.~(\ref{Eq:cutoffpl}) and $G_{1}(E)$ and  
$G_{2}(E)$ by Eq.~(\ref{Eq:gabs}).
The same model is used for all observations to keep consistency between the
parameters from one observation to the next. The last observation 
(revolution 288) is disregarded from this
spectral analysis due to poor statistics.

\begin{figure}
    \centering
    \includegraphics[width=8.7cm]{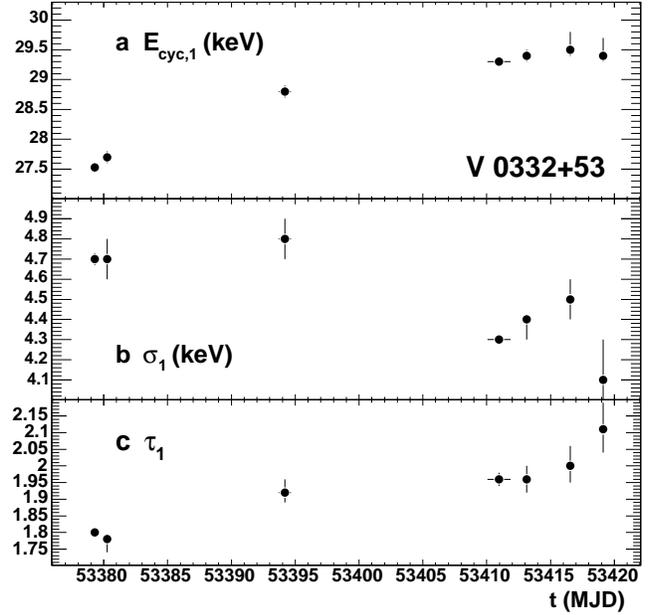}
    \caption{Evolution of the specral fit parameters describing the fundamental
                    cyclotron line. {\bf a} line energy, {\bf b} width and {\bf c} depth.
                    The horizontal bars on the points indicate the duration of each
                    observation.
           }
    \label{Fig:spefit}
\end{figure}

The main fit parameters are summarised in Table~\ref{Tab:spefit}.
The 2--10\,keV and 20--40\,keV fluxes derived from the fitted models confirm
the two phases set forth in Sect.\ref{Sect:flux}, i.e. first an  
exponential decay followed by a linear decrease of the flux.
The folding energy remains about constant around 7.5\,keV, which is  
consistent with the results
of Paper~I (Table~1) for revolution 273--274.
The power law index $\Gamma$, on the other hand, decreases from $-$0.18
in the early revolutions to $-$0.4 in the later ones, in agreement  
with the hardening observed
from the hardness ratios (Fig.~\ref{Fig:hr}).

\begin{figure}
    \centering
    \includegraphics[width=8.7cm]{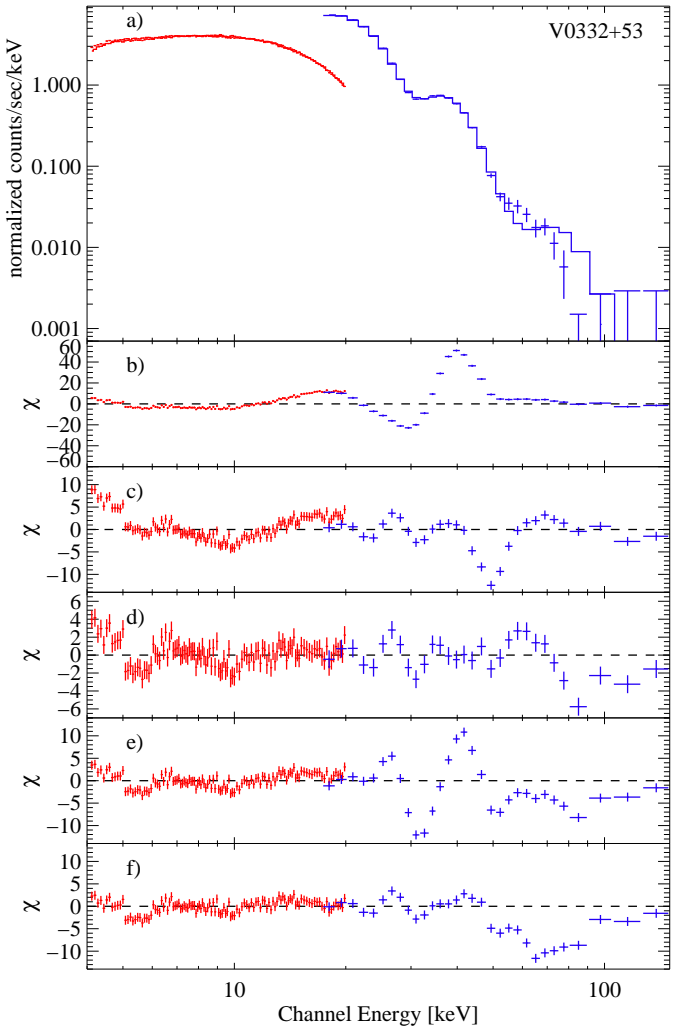}
    \caption{{\bf a} Combined spectrum and folded model of data obtained with JEM-X (left) and ISGRI (right) for revolution 284;
    		{\bf b} residuals of the model with $\Gamma$=$-0.41$ and $E_{\rm fold}$=7.4\,keV, and without any Gaussian line applied;
    		{\bf c} one Gaussian line is included at 29.3\,keV;
    		{\bf d} a second Gaussian is added, at 51\,keV;
		{\bf e} same as {\bf d}, but with the first line fixed at its value of 27.5\,keV found in revolution 273;
		{\bf f} same as {\bf d}, but with $\Gamma$ fixed the value of $-0.17$ found in revolution 273.
           }
    \label{Fig:residuals}
\end{figure}

The cyclotron line parameters for the fundamental are displayed in  
Fig.~\ref{Fig:spefit} as a function of time, and an example of the  
unfolded spectrum with the fit and residuals for revolution 284 is  
shown in Fig.~\ref{Fig:residuals}.
The energy of the fundamental cyclotron line increases by 7\% from
the first to the last \textsl{INTEGRAL} observations, while the depth  
of the line increases by 16\%
and the width decreases by 12\%. The changes are significant;  
attempting to fit the spectrum
in revolution 284 with a line at 27.5\,keV leads to strong residuals  
around the line (Fig.~\ref{Fig:residuals}e). The changes in the line  
parameters cannot be attributed to variations of the continuum  
either; fixing $\Gamma$ in revolution 284 to its value of -0.17 from  
revolution 273 leads to strong residuals at energies above 45 keV  
(Fig.~\ref{Fig:residuals}f).

The determination of the parameters describing the 50\,keV line  
becomes problematic from revolution 284 on, due to poor statistics.
Therefore, for those revolutions, we fix the line energy to the value  
$E_{\rm cycl,2}=51$\,keV obtained for Rev.~273 where the statistics  
are the best, and fit the width and depth of the line to obtain the  
best residals. The results are shown in Table~\ref{Tab:spefit}. The  
width and depth of line are seen to be rather stable throughout the  
whole decay of the outburst, around 9\,keV and 2, respectively. The  
large error bars on the numbers result from the poor statistics.

\subsection{Power spectra}
\label{Sect:powspec}

\begin{figure}
   \centering
   \includegraphics[width=88mm]{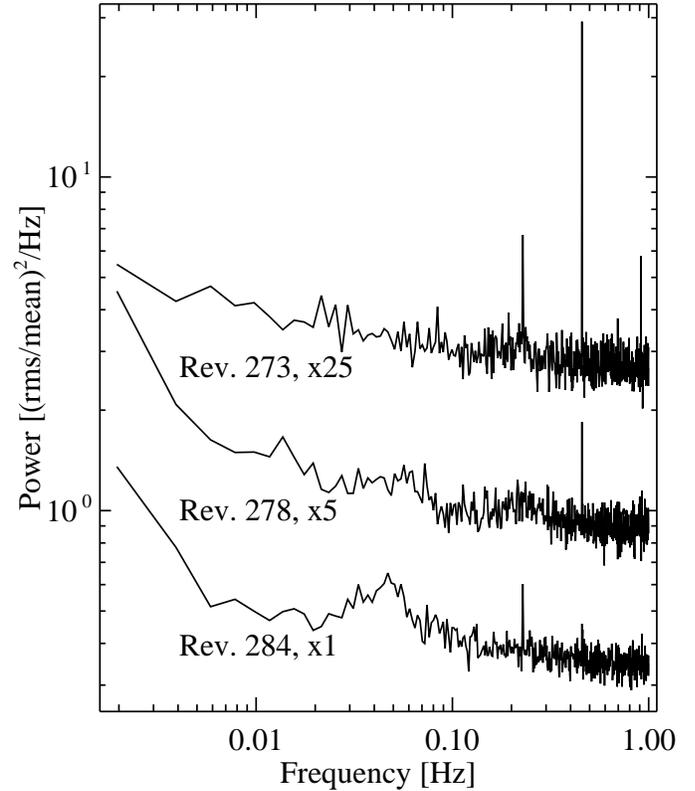}
   \caption{5--20\,keV power spectra, averaged over the indicated
   		revolutions and normalised using the method described by
		Miyamoto et al. (\cite{Miyamoto_etal91}).
		Poisson noise has not been subtracted (see text for more details).
		The power spectra of revolutions 278 and 273
		have been multiplied by factors of 5 and 25, respectively. Error
		bars have been omitted for clarity.}
\label{Fig:powspec_jemx}
\end{figure}

Several X-ray pulsars and cyclotron line sources are known to exhibit
low frequency quasi-periodic oscillations (QPOs) (see, e.g.,
Takeshima et al. \cite{Takeshima_etal91}, Takeshima et al. \cite{Takeshima_etal94},
Heindl et al. \cite{Heindl_etal99}). Takeshima et al. (\cite{Takeshima_etal94})
first discovered a QPO with a centroid frequency of $\sim$0.05\,Hz and
a relative root mean square (rms) amplitude of $\sim$5\% in
2.3--37.2\,keV data from the 1989 outburst of V\,0332$+$53. This
feature has recently also been reported in the 2--60\,keV
\textsl{RXTE}-PCA power spectral density functions (PSDs) obtained
during the 2004 outburst decay (Qu et al. \cite{Qu_etal05}). Furthermore
Qu et al. found a QPO at $\sim$0.22\,Hz, i.e. centered on
the fundamental pulsar frequency, in addition to the narrow peak at
that frequency. The rms strength of this QPO also amounts to a few per
cent. In the following we show that the JEM-X and ISGRI
observations confirm the detection of both features during the
outburst decay, and we analyse their changes with time.
Note that, to our knowledge, these results represent the first QPOs  
reported from \textsl{INTEGRAL} JEM-X and ISGRI data in the literature.

\begin{figure}
   \centering
   \includegraphics[width=88mm]{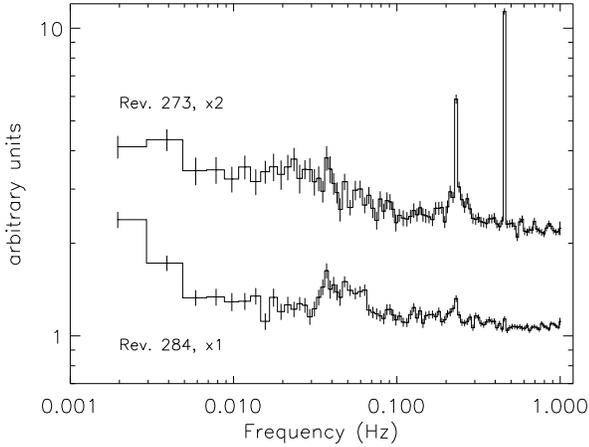}
   \caption{20--40\,keV power spectra, averaged over the indicated revolutions.
   		The spectra are arbitrarily shifted in the y-axis for easy comparison.
		Note that the ISGRI power spectra of revolution 278
		are not shown due to solar flaring giving rise to many gaps in the
		lightcurve after GTI filtering.}
\label{Fig:powspec_isgri}
\end{figure}

The background subtracted 5--20\,keV and 20--40\,keV
lightcurves of each revolution were rebinned to a time
resolution of 0.5\,s and power spectra were produced from 512\,s long
lightcurve segments using Fast Fourier Transforms. Averaged power
spectra for selected revolutions are shown in Figs.~\ref{Fig:powspec_jemx}
and~\ref{Fig:powspec_isgri}.

For coded mask instruments the determination of the white noise level
as well as the influence of the background flux on the power spectrum
can be complex. The white noise component in JEM-X lightcurves,
e.g., is known to be non-Poissonian.
This is also true for the ISGRI which consists of ~16000 pixels, each  
having different noise charactersistics.
Using the normalization described
by Leahy et al. (\cite{Leahy_etal83}) we find the noise level to be  
about one order of
magnitude higher than expected from Poisson noise. However, source
features are well apparent above the noise level, e.g., the 0.05\,Hz
QPO is clearly visible in revolution 284. Therefore, and since the
broad-band PSD (which is not to be discussed here) is dominated by the
noise component up to at least 0.1\,Hz, we did not attempt to correct
the power spectra for white noise. The JEM-X power spectra
shown in Fig.~\ref{Fig:powspec_jemx} are normalised according to
Miyamoto et al. (\cite{Miyamoto_etal91}) and a comparison of the  
strength of features
above the noise level between the JEM-X PSDs and the
quasi-simultaneous PCA PSDs of Qu et al. (\cite{Qu_etal05}) finds  
comparable values
\footnote{This can only be a very rough comparison since different
energy bands are involved and since the power spectrum of V\,0332$+$53
obviously changes rapidly during the decay.}.

Strong changes in the shape of the power spectrum can be seen over the
outburst decay, where the varying amplitude ratios at the pulsar
frequency and its harmonics reflect a strong change of the pulse
profile (Zhang et al. \cite{Zhang_etal05}).
The PSDs shown in Figs.~\ref{Fig:powspec_jemx}
and~\ref{Fig:powspec_isgri} represent the main stages of the
evolution.
Although, for the reasons discussed earlier, the white noise level is  
not accurately estimated in the INTEGRAL data, the JEM-X and ISGRI  
PSDs for revolutions 0273 and 0284 do compare well and show similar  
features as a function of frequency.

In revolution 273 (Jan. 8--9) the power spectrum is dominated by
signatures of the pulsation frequency, especially a very strong first
harmonic. In addition weak red noise is present. The peak at the
fundamental pulsation frequency is slightly broadened at the bottom,
especially in the ISGRI PSD. However, the 0.22\,Hz QPO is not as
obvious as a few days later in the Jan. 15--19 PCA power spectrum of
Qu et al. (\cite{Qu_etal05}). In revolution 278 (Jan. 23--24) the fundamental
pulsation at 0.22\,Hz is not seen anymore but a weak broad QPO feature
is present at this frequency. The first harmonic is still clearly
visible. The strength of the red noise component more than doubled at
low frequencies and there is an indication of additional power in the
frequency region of the 0.05\,Hz QPO. Principally the overall PSD
shape is very similar to the Jan. 15--19 PCA power spectrum mentioned
above. Finally, in revolution 284 (Feb. 9--10) the 0.05\,Hz QPO is
clearly visible, especially in the JEM-X power spectrum. The
red noise is still strong while the pulsation fundamental is back as a
narrow feature, stronger than the first harmonic. Fitting the low
frequency QPO in revolution 284 with a Lorentzian gives ($0.049\pm0.001$)\,Hz, with a FWHM of ($0.023\pm 0.005$)\,Hz for JEM-X
and ($0.049\pm 0.003$)\,Hz with a FWHM of ($0.021\pm 0.008$)\,Hz for
ISGRI, where the underlying continuum can be modelled with a flat power
law component and a very low frequency Lorentzian.

\section{Discussion}
\label{Sect:discussion}

V\,0332+53 is a highly magnetised pulsar which has a magnetic field  
of about $2.7\times 10^{12}$\,G (Paper~I, Pottschmidt et al. \cite{Pottschmidt_etal05}) and orbits a massive O8-9Ve companion with a  
period of about 34 days.
The X-ray outbursts from this system are observed with a time recurrence of about 10 years.
They are believed to be triggered by a sudden increase of the mass ejection rate from the massive star,
the signature of which has been observed in the optical brightening  
of the system in 2004 (Goranskij \& Barsukova \cite{Goranskij_Barsukova04}).
The ejected mass feeds an accretion disk around the pulsar, from  
which it is funneled by the magnetic field towards the polar caps and  
falls on the surface of the neutron star.

The analysis of the 2005 outburst presented in Sect.~\ref{Sect:results} provides
new hints towards a better understanding of V\,0332+53 within this basic picture.
The X-ray lightcurve (Sect.~\ref{Sect:flux}) displays a  
characteristic exponential decay with a time scale of 20 to 30 days,
followed by a linear decrease. This feature may provide some  
information on the disk properties.
The X-ray spectrum (Sect.~\ref{Sect:spectrum}) is well fitted by a  
folding power law modified by two cyclotron lines.
The 27\,keV energy of the fundamental cyclotron line is shown to  
increase by 7\% during the outburst decay
and to become narrower by ~12\%. At the same time, its depth  
increases by 16\%.
This information would provide some insight into the resonance  
scattering region near the polar caps.
Finally, the appearance of the 0.05\,Hz QPO only from the middle of  
the decay on (Sect.~\ref{Sect:powspec}) provides further information  
on the properties of the matter around the neutron star.
Each of those observational facts are successively discussed in the  
following sections.

\subsection{Lightcurve}
\label{Sect:discussion:lightcurve}

An exponential decay of the flux is observed in dwarf novae and in  
soft X-ray transients
on timescales from a few days for the former to several weeks for the  
latter
(e.g. Lasota \cite{Lasota96}, Tanaka \& Shibazaki \cite{Tanaka_Shibazaki96}).
King \& Ritter (\cite{King_Ritter98}) show how illuminated disks in  
low-mass X-ray binary stars
can produce such an exponential decay on the time scales observed for  
the soft X-ray transients.
In that picture, the central star illuminates the disk and keeps its  
temperature high.
The disk empties proportionally to its mass, leading to an  
exponential decay of the emitted flux.
They further show that the lightcurve would be characterised by a  
linear decrease at later times.
The change to a linear regime is triggered by a change in the  
conditions in the disk.
In low-mass X-ray binaries the change corresponds to the decrease of  
the disk temperature
below the temperature of partial H ionisation, such that the high  
temperature
cannot hold over the entire disk anymore.
This scenario of X-ray irradiated accretion disks has subsequently  
been confirmed by numerical calculations (Dubus et al. \cite{Dubus_etal99}, \cite{Dubus_etal01}).

It is tempting to make the parallel with the results found for V\,0332+53.
The source displays an exponentially decreasing flux followed by a  
linear decrease, similar to the disk-irradiated systems.
This is the first time, to our knowledge, that such a behaviour is  
observed in a high-mass X-ray binary.
Since the X-ray flux is proportional to the mass accretion rate, this  
picture suggests that
the accretion rate is proportional to the mass of the disk during the  
exponential decay phase.
The transition to a linear decrease phase would be triggered by an as  
yet un-identified change of conditions in the disk.

The two different exponential decay times observed in the lightcurves  
of V\,0332+53, of ~30 days above 20\,keV and ~20 days below (Sect.~\ref{Sect:flux}) suggest the presence of at least two regions  
contributing to the X-ray continuum of that source.
The harder component would originate from the accretion column
near the resonant scattering region, while the softer component  
would originate from a region located, for example, higher in  
the accretion column.
The observed increase of the depth of the cyclotron line with time  
is also compatible with this scenario; the softer component  
decays more rapidly and hence its tail would contribute relatively  
less to the high energy part of the spectrum from which the photons  
are scattered by the electrons in the magnetic field.
Basically, one time scale governs the rate of mass flow onto the  
neutron star through the disk, leading to the exponential decay  
followed by a linear decrease (see above); another time scale governs  
the spectral changes of the emission that may come from different  
heights within the accretion column.

\subsection{Spectrum}
\label{Sect:discussion:spectrum}

Time evolution of cyclotron resonance line parameters has already  
been reported in the literature for some binary systems.
Mihara et al. (\cite{Mihara_etal98}) report a relation between the  
line energy and the X-ray luminosity for 5 pulsars observed by {\it Ginga}.
The resonance energy of 4U~0115+53, for example, increased by 40\% as  
the luminosity decreased by a factor of 6 between 1990 and 1991. The  
1990 outburst of V~0332+53 is also reported in that paper, with a 10\%
decrease of the line energy from 30.0$\pm$0.5\,keV to 27.2$\pm$0.3 
\,keV as the {\it Ginga} count rate in the 3-37\,keV band increased  
by a factor of 1.6.
These authors attribute the change of the line energy to that of the  
height of the scattering region in a dipole magnetic field, leading  
to a height change of $\sim$1.1\,km for 4U~0115+53 and $\sim$330\,m  
for V~0332+53.
Kreykenbohm et al. (\cite{Kreykenbohm_etal04}), on the other hand,  
report a variation of the cyclotron energy of GX\,301-2 with the  
pulse phase of the pulsar, and attribute the change to different  
viewing angles of the accretion column where the line originates.

In the case of the 2005 outburst decay of V\,0332+53 analysed in this  
paper, the increase with time of the energy and depth of the  
cyclotron fundamental line, together with the decrease of its width  
can also be understood as resulting from the variation of the  
scattering region characteristics in a dipole magnetic field above  
the polar caps of the pulsar.
It is interesting to note that Pottschmidt et al. (\cite{Pottschmidt_etal05}), analyzing the {\it RXTE} data of V\,0332+53  
taken two weeks before the first \textsl{INTEGRAL} observation,  
derive E$_{1}=25.2$\,keV for 2004 December 24--26, which confirms the  
trend of increasing line energy with time found in this paper.
Let us thus assume that the region where the cyclotron resonance  
scattering takes place extends over some fraction of the
accretion column. Then the region would cover some range of magnetic strengths,
and a broad line shape would result from the superposition of a succession of narrow lines,
each originating from a different height and magnetic field strength in the accretion column.
As the accretion rate and hence the flux decreases with time, the extension of the
resonance scattering region is expected to decrease along with the total width of the line.
At the end of the decline only the region close to the neutron star surface, where the magnetic
field is stronger, would contribute to the resonance scattering; this  
would lead to a narrower line shape at a higher energy.
The observed energy E$_{1}$ indeed increases by 17\% from the RXTE  
observation of December 2004 to the last \textsl{INTEGRAL} observation in February 2005.
If we consider a dipole approximation for the magnetic field, we have  
$B\sim r^{-3}$ in a non-relativistic description, where $r$ is the  
radius to the center of the neutron star.
The extent of the resonance scattering region at the peak of the  
outburst would then, in this first scenario, be about 6\% of the  
stellar radius, i.e. about 500 meters for typical radii of neutron  
stars.  A change in the scattering energy between two different  
observations of Her~X-1, from 34\,keV to 41\,keV, has been discussed  
by Gruber et al. (\cite{Gruber_etal01}); they conclude that the range  
over which the scattering occurs is of the order of 1 km in height,  
if their observed line width of 5 keV is largely due to a mix of  
magnetic field strengths in a dipole field.  In the case of V\,0332+53,
we additionally observe an evolution of the width and depth of  
the line with time.

Alternatively, we can argue that the fundamental line originates from  
a unique narrow region in the column density, but that the line shape  
is quite complex and variable with time, resulting in variations of  
the line width and depth.
An increase in the line energy would then be related to a  
displacement with time of the scattering region.
In the case of V~0332+53, the observed 17\% increase of E$_{1}$ would  
imply a move of the location of the scattering region by 500 meters  
in the accretion column down to the surface of the neutron star.
However, in this second scenario, the line profile is expected to  
remain approximately constant with time, which is not supported by  
the \textsl{INTEGRAL} observations.
Also, if the scattering region gets closer to the neutron star,  
higher order terms in the B-field may increase the line width, which  
is opposite to the trend observed by \textsl{INTEGRAL}.
We therefore would rather support the first scenario.

\subsection{Quasi periodic oscillations}
\label{Sect:discussion:QPO}

The 0.22\,Hz QPO observed around the spin frequency in both the \textsl{INTEGRAL}
(Sect.~\ref{Sect:powspec}) and RXTE (Qu et al. \cite{Qu_etal05}) data must originate
from a region around the neutron star co-rotating with the pulsar.
This region can either be located at the inner radius of the  
accretion disk near the Alfven radius, or in the corona above the hot  
spots, or somewhere in between along the magnetic field through which  
the matter is tunneled.

The origin of the lower frequency QPO at 0.05\,Hz, on the other hand,  
is more debatable.
Low frequency QPO-type variabilities are known in several X-ray  
pulsars (see van der Klis \cite{van_der_Klis04} for a review).
Although there is no general consensus on the origin of QPOs, most models do invoke the
presence of some sort of inhomogeneities in the accretion disk.
With a typical neutron star mass $M_{\rm NS}$=1.4\,${\rm M_\odot}$  
and a spin of  $P_{\rm NS}\simeq 4$\,s for V\,0332+53,
the co-rotation radius where the angular velocity of the  
magnetosphere and the Keplerian velocity of the disk are equal is
\begin{equation}
   R_{\rm corot} \simeq 1500 \, \left(\frac{M_{\rm NS}}{M_{\odot}}\right)^{1/3}
                                                                    \left(\frac{P_{\rm NS}}{sec}\right)^{2/3}               \; {\rm km}
                           \;\simeq\; 4230 \, {\rm km}.
\end{equation}
On the other hand, the radius at which the Keplerian frequency is equal to that of the 0.05 
\,Hz QPO is $\sim$12350\,km.
The scenario of having the 0.05\,Hz QPO somehow related to the  
part of the disk at this distance, i.e. three times further away  
than the co-rotation radius, is thus plausible.
The temperature of the disk cannot be responsible for the observed  
emission, since no black body emission is found in the spectrum.
The QPO could result from an occultation of the beam by intervening  
matter from the disk at that distance, as suggested by Heindl et al.  
(\cite{Heindl_etal99}) to explain a 2\,mHz QPO observed in 4U\,0115+63.


\begin{thebibliography}{}
\bibitem[2005] {Coburn_etal05} 				Coburn W., Kretschmar P., Kreykenbohm I., et al. 2005, ATel 381, 1
\bibitem[1999] {Dubus_etal99}					Dubus G., Lasota J.-P., Hameury J.-M., Charles P. 1999, MNRAS 303, 139
\bibitem[2001] {Dubus_etal01}					Dubus G., Hameury J.-M., Lasota J.-P. 2001, A\&A 373, 251
\bibitem[2004] {Goranskij_Barsukova04}	Goranskij V., Barsukova E. 2004, ATel, 245, 1
\bibitem[2001] {Gruber_etal01}					Gruber D.E., Heindl W.A., Rothschild R.E., Coburn W., Staubert R., Kreykenbohm I., Wilms J. 2001, ApJ 562, 499
\bibitem[1999] {Heindl_etal99}					Heindl W.A., Coburn W., Gruber D.E., etal. 1999, ApJ 521, L49
\bibitem[1998] {King_Ritter98}					King A.R., Ritter H. 1998, MNRAS 293, L42
\bibitem[1999] {Kreykenbohm_etal99} 		Kreykenbohm I., Kretschmar P., Wilms J., Staubert R., Kendziorra E., Gruber D. E., Heindl W. A., Rothschild R. E. 1999, A\&A 341, 141
\bibitem[2004] {Kreykenbohm_etal04} 		Kreykenbohm I., Wilms J., Coburn W., Kuster M., Rothschild R. E., Heindl W. A., Kretschmar P., Staubert R. 2004, A\&A 427, 975
\bibitem[2005] {Kreykenbohm_etal05} 		Kreykenbohm I., Mowlavi N., Produit N., et al. 2005, A\&A 433, L45 (Paper I)
\bibitem[1996] {Lasota96} 						Lasota J.P. 1996, IAUS 163, 43
\bibitem[1983] {Leahy_etal83}					Leahy D.A., Darbro W., Elsner R.F., et al. 1983, ApJ 266, 160
\bibitem[1996] {Levine_etal96} 				Levine A.M., Bradt H., Cui W., Jernigan J. G., Morgan E.H., Remillard R., Shirey R.E., Smith D.A. 1996, ApJ 469, L33
\bibitem[1998] {Mihara_etal98} 				Mihara T., Makishima K., Nagase F. 1998, Adv. Space Res. 22, 987
\bibitem[1991] {Miyamoto_etal91} 			Miyamoto S., Kimura K., Kitamoto S., et al. 1991, ApJ 383, 784
\bibitem[2005] {Pottschmidt_etal05}			Pottschmidt K., Kreykenbohm I., Wilms J., Coburn W., Rothschild R.E.,
																Kretschmar P., McBride V., Suchy S., Staubert R. 2005, ApJ 634, L97
\bibitem[2005]{Qu_etal05}			Qu J., Zhang S., Song L., Falanga M. 2005 ApJ 629, L33
\bibitem[2004] {Swank_etal04}					Swank J., Remillard R., Smith E. 2004, ATel 349
\bibitem[1996] {Tanaka_Shibazaki96}		Tanaka Y., Shibazaki N. 1996, ARAA 34, 607
\bibitem[1991] {Takeshima_etal91}			Takeshima T., Dotani T., Mitsuda K., Nagase F. 1991, PASJ 43, L43
\bibitem[1994] {Takeshima_etal94}			Takeshima T., Dotani T., Mitsuda K., Nagase F. 1994, ApJ 436, 871
\bibitem[2004]{van_der_Klis04}				van der Klis, M. 2004, in ñCompact stellar X-ray sourcesî, Lewin \& van der Klis (eds.), Cambridge University Press [astro-ph/0410551]
\bibitem[2003] {Winkler_etal03}				Winkler C., Courvoisier T. J.-L., Di Cocco G. et al. 2003, A\&A 411, L1
\bibitem[2005] {Zhang_etal05}					Zhang S., Qu J.L., Song L.M., Torres D.F. 2005, ApJ 630, L65
\end{thebibliography}
\end{document}